\documentclass[11pt,a4paper]{article}
\usepackage{ifpdf}
\usepackage{ae}
\usepackage[T1]{fontenc}
\usepackage[ansinew]{inputenc}
\usepackage{mathrsfs}
\usepackage{amsmath}
\usepackage{amssymb}
\pdfoutput=1
\usepackage{amsmath,amssymb,amsfonts,a4wide,graphicx,bm,times,psfrag,wrapfig,sidecap}
\usepackage{cite}
\usepackage[colorlinks=true,linkcolor=black, citecolor=black,
urlcolor=black]{hyperref} 
\numberwithin{equation}{section}
\makeatletter \let\old@startsection=\@startsection
\renewcommand{\@startsection}[6]
{\old@startsection{#1}{#2}{#3}{#4}{#5}{#6\mathversion{bold}}}
\makeatother

\newcommand\re[1]{({\ref{#1}})}
\def\be{\begin{eqnarray}  }
    \def\ee{\end{eqnarray}}
 
 \def\IR{{\mathbb{R}}}
\def\IZ{{\mathbb{Z}}}
 
 \def\IN{{\mathbb{N}}}

    \def\la{\label}

    \def\({\left(} \def\){\right)} \def\<{\langle\,} 
    \def\>{\,  \rangle} \def\[{\left[} \def\]{\right]} 
        
      \def\CO{{ \mathcal{ O} }}
     \def\CZ{{ \mathcal{ Z} }}  
            \def\CF{{\cal F}}
     \def\p{\partial} \def\a{\alpha} 
    \def\g{\gamma} \def\e{\epsilon}  
      \def\G{\Gamma} 
  
 \def\Tr{{\rm Tr}}

\def\d{\delta}

\usepackage[ansinew]{inputenc}
\newcommand{\imineq}[2]{\vcenter{\hbox{\includegraphics[height=#2ex]{#1}}}}
\def\II{{\mathbb{I}}}
   \def\tv{{\tilde v}}
  \def\uu{{\bf u}}
  \def\tphi{{\tilde\phi}}
 \def\tp{{\tilde p}}
 \def\tS{{\tilde S}}
 \def\tE{{\tilde E}}

\begin{document}

\newcommand\encadremath[1]{\vbox{\hrule\hbox{\vrule\kern8pt
\vbox{\kern8pt \hbox{$\displaystyle #1$}\kern8pt}
\kern8pt\vrule}\hrule}} \def\enca#1{\vbox{\hrule\hbox{
\vrule\kern8pt\vbox{\kern8pt \hbox{$\displaystyle #1$} \kern8pt}
\kern8pt\vrule}\hrule}}


\thispagestyle{empty}

\begin{flushright}
\end{flushright}

\vspace{1cm}
\setcounter{footnote}{0}

\begin{center}

 {\Large\bf TBA  and tree expansion }

\vspace{20mm} 

Ivan Kostov\footnote{\it Associate member of the Institute for
Nuclear Research and Nuclear Energy, Bulgarian Academy of Sciences, 72
Tsarigradsko Chauss\'ee, 1784 Sofia, Bulgaria},
  Didina Serban and Dinh-Long Vu \\[7mm]
 
{\it  Institut de Physique Th\'eorique, CNRS-UMR 3681-INP,
 C.E.A.-Saclay, 
 \\
 F-91191 Gif-sur-Yvette, France}, 
 \\[5mm]

\end{center}

\vskip9mm

\vskip18mm

{ 
 \noindent{ We propose an alternative, statistical, derivation of the
Thermodynamic Bethe Ansatz  based on the tree expansion of the 
Gaudin determinant.  We
illustrate the method on the simplest example of a theory with
diagonal scattering and no bound states.  We reproduce the expression
for the free energy density and the finite size corrections to the
energy of an excited state as well as the LeClair-Mussardo series for
the one-point function for local operators.  }
 }

\newpage
\setcounter{footnote}{0}

\section{Introduction}
\label{sec:1}

The finite size effects in 1+1 dimensional field theories come from
the quantisation of the momenta of the physical particles, as well as
from the virtual ``mirror'' particles winding around the space circle
$R$ \cite{Luscher:1983rk}.  When $R$ is large, the exponentially small
contribution from the mirror particles can be neglected and the
spectrum is determined by the ``asymptotic'' Bethe-Yang equations,
which take into account only the scattering processes between the
physical particles.  As it was first realised by Al.  Zamolodchikov
\cite{Zamolodchikov:1989cf}, for finite $R$ a powerful technique for
summing up the finite size corrections is given by the Thermodynamical
Bethe Ansatz, or TBA \cite{YY}.  If the the theory is Lorentz
invariant, the finite size effects can be traded to finite temperature
effects.  The main idea of the TBA is that the thermal trace is
dominated by a saddle point for the density of states, which is
obtained as the solution of some non-linear integral equations.  By
analytical continuation one can obtain the ``exact Bethe equations''
for the spectrum of the excited states in finite volume
\cite{Dorey:1996re}.
 
In the last decades much attention is been focused on combining the
TBA and the form factor bootstrap in order to compute the correlation
functions at finite volume/temperature.  This is a problem of higher
complexity and in spite of the considerable progress a systematic
procedure is not yet available for the higher point functions.  The
main difficulty is to learn how to insert efficiently the resolution
of the identity between the local operators in order to split the
correlation function into simpler objects, the elementary form factors
at infinite volume.  In other words, the saddle point analysis of the
TBA is not sufficient and has to be replaced by a more subtle,
field-theoretical, consideration.
     
Another motivation for looking at the sum over the intermediate states
is the recently proposed hexagon bootstrap program in the AdS/CFT
integrable model \cite{BKV1} which can be applied for the computation
of higher point correlation functions.  The proposal prescribes to
 insert  complete sets of mirror particles
 between  the hexagon operators.  Although these effects 
resemble the wrapping corrections in the spectral problem,   no TBA methods 
have yet been developed to resum them.
  
In this paper, we address the problem of performing the sum over the
mirror states in the simplest case of a theory with diagonal
scattering and no bound states.  Our proposal is close in spirit to
some previous works \cite{Balog-TBA}\cite{Woynarovich:2010wt} where
the excluded volume in the sum over the intermediate states is
compensated by including into the sum non-physical solutions of the
asymptotic Bethe-Yang equations.  The new development is that we
succeeded to perform explicitly the sum over the states using a graph
expansion of the Gaudin determinant  which gives the integration
measure over the Bethe states in the mirror channel.
 This graph
expansion leads to a Feynman-like diagram technique which allows us to
write the free energy as a sum over tree Feynman diagrams.
 
In section \ref{section:PartFun} we explain our method on the simplest
example of a diagonal theory without bound states for which we compute
the the partition function on a cylinder with circumference $R$ as the
thermal trace in the mirror theory.  In the rest of the text we
consider two more examples, where we re-derive the formulas obtained
previously by ingenuous application of the TBA. In section
\ref{section:Casimir} we compute the energy of an excited state in the
physical channel.  In section \ref{section:LM} we derive the
LeClair-Mussardo series for the one-point function.  In all three
examples we reduce the computation to a combinatorial problem
involving the sum over tree graphs.

  \section{ Integrable Quantum Field Theory on a cylinder: the
  partition function}
  \label{section:PartFun}
  
  \subsection{Physical and mirror channels}

Consider an integrable 1+1 dimensional field theory with one single
type of particle excitations above the vacuum.  The dispersion
relation between the momentum $p$ and the energy $E$ of the particle
is parametrised by the rapidity variable $u$:
 \be p=p(u), \ E= E(u).  \ee
We assume that there exists a transformation to the ``mirror'' theory
in which the role of the time $t$ and the space $x$ are exchanged.
The physical and the mirror channels are related by a ``mirror''
transformation $x=-i\tilde t, t = -i \tilde x$ and $E= i \tilde p, p =
i \tilde E$.  The mirror transformation can be encoded in a
transformation $\g: u\to \tilde u$ of the rapidity parameter, so that
\be E(\tilde u )= i \tilde p(u), \quad p(\tilde u)= i \tilde E(u).
\ee
The square of the mirror transformation gives the crossing
transformation $\g^2: u\to \bar u= \g \tilde u $ which relates
particles to anti-particles.  If the theory is Lorentz invariant, then
the mirror and the physical theories are identical.  The diagonal
S-matrix $S(u, v)$ is supposed to satisfy, besides the Yang-Baxter
equations, unitarity $S(u, v)S(v,u)=1$, crossing symmetry $S(u, v) =
S(\bar v, \bar u)$, and the condition $S(u,u)=-1$.  We will not need
to assume that the S-matrix is a function of the difference of the two
rapidities.
   
If the theory is confined in a finite volume $R$ with periodic
boundary conditions, the eigenstates of the Hamiltonian can be
constructed as superpositions of plane waves according to the Bethe
Ansatz, with the spectrum of the rapidities determined by condition of
periodicity.  Each eigenstate from the $N$-particle sector is
characterised by a set of rapidities $\uu= \{u_1, \dots, u_N\}$ and
the energy of this state is
  equal to %
  \be \la{physenergy} E(\uu)= \sum_{j=1}^N E(u_j).  \ee
 When $R$ is sufficiently large, the spectrum of the energies are
 determined by the Asymptotic Bethe Ansatz.  The quantisation
 condition for the rapidities is expressed in terms of the total phase
 factor corresponding to a process in which one of the $N$ particles
 winds once around the space circle,
\be
\label{Betheq}
  \phi_j (u_1, \dots , u_N)\equiv p(u_j)R + {1\over i}\sum_{k(\ne
  j)}^N \log S(u_j,u_k) \qquad (j=1, \dots, N).  \ee
 For periodic boundary conditions the scattering phases can take
 integer values modulo $2\pi$
   \be \la{ABA} \phi_j(u_1,\dots, u_N)= 2\pi n_j \ \ \ \mathrm{with} \
   n_j \ \mathrm{ integer}, \quad j=1, \dots, N\, .  \ee
In a system of units where the mass of the particle is equal to one,
the asymptotic expression \re{Betheqb} for the scattering phases is
true up to $o(e^{-R})$ terms.  For finite $R$ the Bethe-Yang equations
\re{Betheq}-\re{ABA} are deformed by the scattering with the virtual
particles in the mirror channel which wrap the space circle \cite{Luscher:1985dn}.  
One can study the finite volume effects using the TBA in the mirror channel.
One can introduce an infrared cutoff in the mirror theory by
considering the cylinder as the limit of a torus obtained as the
product of the space circle with a time circle with asymptotically
large circumference $L$.  When $L$ is large, one can construct a
complete set of states in the mirror channel whose spectrum is given
by the asymptotic Bethe-Yang equations.  Then the partition function
can be computed by taking the thermal trace in the mirror Hilbert
space.

The standard TBA approach due to Yang and Yang \cite{YY} is to express
the thermal trace as an integral over the density of one-particle
rapidities, taking into account both the energy and the entropy of the
states.  The free energy is expressed as a functional of the rapidity
density and the critical point of this functional gives both the
thermal equilibrium state and the expression for the extensive piece
$L F_0(R) $ of the free energy.  In field-theoretical terms this
translates to replace the sum over the intermediate states by a single
``thermal state'' characterised by the saddle point density.  This
approximation works well for evaluating the free energy and the
one-point functions, where a single insertion of the identity is to be
made, but it is not sufficient e.g. for the computation of the
two-point functions.\footnote{There however is a class of two-point
functions for which a single insertion is sufficient
\cite{Pozsgay:2018eki}.}

 \subsection{ Thermal partition function}

\label{sec:2}

Below we will perform a direct summation in the mirror Hilbert space.
Our method is exact up to corrections exponentially small in $L$ and
allows to control the whole $1/L$ expansion of the partition function.
The simplest object to compute is the partition function on the torus,
$Z(R,L)$, which can be evaluated as a thermal trace in the physical or
in the mirror channels of the Euclidean theory,
\be \la{ThermalTrace} \CZ(L,R)= \underset{\text{phys}}\Tr [e^{- L
H_{\text{phys}} }] = \underset{\text{mir}}\Tr [e^{- R H_{\text{mir}}
}].  \ee
Assuming that $R \ll  L$, our goal is to evaluate the the free
energy
\be \la{FREEXP} \log \CZ(L,R)= L F_0(R)+ F_1(R) + \dots \ee
up to corrections exponentially small in $L$.

Let us stress that such an exponential accuracy is beyond the reach of
the standard TBA approach which is essentially a collective field
theory for the rapidity density and as such suffers from ambiguities
beyond the first two terms of the expansion \re{FREEXP}.  The leading
term in the TBA approach is determined by the saddle point of the
integral over the densities, while the subleading term is produced by
the gaussian fluctuations about the saddle point
\cite{Woynarovich:2004gc} and the normalisation of the wave function
of the thermal state \cite{Pozsgay:2010tv}, with the two effects
cancelling completely for periodic boundary conditions.  Our approach
does not suffer from the ambiguities of the collective theory and
allows to obtain the whole series \re{FREEXP}, which in the case of
periodic boundary conditions consists of a single term $LF_0(R)$.

  \subsection{The partition function as a sum  over mode numbers}
  \label{subsection:modenumbers}

 The quantisation condition in the mirror channel is given by the
 Bethe-Yang equations
\be \la{ABAb} \tphi_j= 2\pi n_j \ \ \ \mathrm{with} \ n_j \ \mathrm{
integer}, \quad j=1, \dots, M\, , \ee
  where $\tphi _j$ is the total scattering phase for the $j$-th mirror
  particle,
\be
\label{Betheqd}
\tphi_j (u_1, \dots , u_M)\equiv \tp(u_j)L + {1\over i}\sum_{k(\ne
j)}^M \log \tS(u_j,u_k).  \ee
 Here $\tilde S(u,v) = S(\tilde u, \tilde v)$ denotes the S-matrix for
 the mirror particles.  The states in the $M$-particle sector of the
 Hilbert space are labeled by $M$ distinct mode numbers $n_1, \dots,
 n_M$ and the identity operator in this sector can be decomposed as a
 sum of products of normalised states
\be \la{expiden} \II_M= \sum_{n_1<...< n_M} | n_1,\dots, n_M\rangle \langle
n_1,\dots, n_M| .  \ee
If we denote by $\tE_M(n_1, \dots, n_M)$ the eigenvalue of the
Hamiltonian for the state $|n_1, \dots, n_M\rangle$, the partition
function \re{ThermalTrace} is given by the series
\be \la{PrtfnDisc} \CZ(L , R)= \sum_{M=0}^\infty \ \
\sum_{n_1<n_2<\dots <n_M} e^{- R\tE(n_1,\dots, n_M)} .  \ee

 Our goal is to replace in the thermodynamical limit $L\to \infty$
 the discrete sums by multiple integrals.  For that we have first to
 get rid of the ordering of the quantum numbers.  For that we insert a
 factor which kills the configurations with coinciding quantum numbers
 and take the sum over non-restricted integers,
  \be \la{Prtfnonordered} \la{PrtfnDiscb} \CZ(L,R)=\sum_{M=0}^\infty \
  \ {1\over M!} \sum_{n_1 , \dots, n_M} \ \prod_{j<k}^M\(1- \d_{n_j,
  n_k}\) e^{- R \tE(n_1,\dots, n_M)} .  \ee
 Expanding the product of Kronecker symbols, leads to a series
 \be
 \la{cumexp}
 \begin{split}
  \CZ(L ,R)&= 1+ \sum_n e^{-R\tE(n)} + {1\over 2!} \sum_{n_1, n_2}
  e^{-R\tE(n_1,n_2)} - {1\over 2} \sum_{n} e^{- R\tE(n,n)}+ \dots
 \end{split}
 \ee
 which we are going to write as an exponential.  The sum in
 \re{cumexp} goes over all sequences $(n_1^{r_1} , \dots , n_m ^{r_m}
 )$ of positive integers $n_j$ with multiplicities $r_j$.  For
 exemple, $(n^2)= (n,n)$.  Each such sequence defines an (unphysical)
 Bethe state obtained by identifying some of the momenta of a Bethe
 state with $M=r_1+\dots + r_m $ magnons.  This state is a linear
 combination of plane waves with momenta $r_j \tp(u_j)$, $j=1, \dots,
 m$.  and energy
 \be \la{interactionenergy} \tE(n_1 ^{r_1}, \ \dots\ , \, n_m ^{r_m})
 =r_1 \tE(u_1) +\dots + r_m \tE(u_m).  \ee
The relevance of such states has been already pointed out by
Woynarovich \cite{Woynarovich:2010wt} and by Dorey {\it et al }
in

.  The rapidities $u_1, \dots,
u_m$ are determined by the Bethe-Yang equations \re{ABAb} with
$M=r_1+\dots + r_m $.  The phase $\tilde \phi_j$ is acquired by the
wave function if to one of the $r_j$ particles with rapidity $u_j$
winds once around the time circle,
\be
\label{Betheqg}
   \tphi_j \equiv \tp(u_j)L + {1\over i}\sum_{k(\ne j)}^m r_k\log
   \tS(u_j,u_k) + \pi (r_j-1) = 2\pi n_j \quad (j=1, \dots, m ).  \ee
The term $\pi (r_j-1) $ originates in the scattering of the probe
particle with the $r_j-1$ particles with the same rapidity $u_j$.

The full series \re{cumexp} has the form
 \be
 \la{partFFC}
 \begin{split}
\CZ(L ,R) &= \sum_{m=0}^\infty {(-1)^m \over m!} 
 \sum_{n_1, \dots, n_m} \sum_{r_1, \dots,
r_m} (-1)^{r_1+\dots +r_m} \ C_{ r_1\dots r_m} \ e^{- R \tE(n_1
^{r_1}, \ \dots\ , \, n_m ^{r_m})} ,
     \end{split}
 \ee
 where the coefficients $ C_{ r_1\dots r_m}$ are purely combinatorial.
 They can be fixed from the expansion of the thermal partition
 function when the quasiparticles are free fermions, $S(u_i,u_j)=-1$
 and $\tE(n_1, \dots, n_M)= \tilde E(n_1)+\dots + \tilde E(n_M)$.  In
 the occupation numbers representation, the partition function for
 free fermions can be written as an infinite product
 \be
 \la{partFF}
 \begin{split}
    \CZ ^{\text{free fermions}} &= \prod_{n\in\IZ}\(1+ e^{-R
    \tE(n)}\)=\exp \sum_{n\in\IZ} \, \sum_{r=1}^\infty {
    (-1)^{r-1}\over r} e^{- r R \tE(n)} \\
 &= 1+ \sum_{m=1}^\infty {(-1)^m \over m!} \sum_{n_1, \dots, n_m}
 \sum_{r_1, \dots, r_m} {(-1)^{r_1+\dots+ r_m}\over r_1\dots r_m}
 \prod_{j=1}^m e^{- R r_j E(n_j)} .
     \end{split}
 \ee
Comparing with \re{partFFC} we find for the combinatorial coefficients
  \be \la{Ccoefs} C_{r_1\dots r_m}= {1\over r_1\dots r_m}.  \ee

In the case of free fermions, the multiplicities $r_j$ have obvious
meaning.  The vacuum energy is a sum of all fermionic loops including
those winding $r$ times around the space circle.  The weight of an
$r$-winding loop consists of a Boltzmann factor $e^{- r R E_n}$, a
sign $(-1)^r$ due to the Fermi statistics and a combinatorial factor
$1/r$ counting for the $Z_r$ cyclic symmetry.  It is natural to
interpret the multiplicities $r_j$ as winding, or wrapping, numbers
also in the case of non-trivial scattering, which we are going to do
in the following.

  \subsection{ From   mode numbers to  rapidities}
  \la{subsection:mntorap}

The discrete sum over the allowed values of the phases $\tphi
_j(u_1,r_1; \dots, u_m,r_m)$ for given wrapping numbers can be
replaced, up to exponentially small in $L$ terms, by an integral,
\be \la{intphi} \sum_{n_1, \dots, n_m} =\int {d\tphi _1\over 2\pi}
\dots {d\tphi _m\over 2\pi } .  \ee
Since the energy takes a simple form as a function of the rapidities,
eq.  \re{interactionenergy}, we are going to change the variables from
scattering phases $\phi_j$ to rapidities $u_j$,
 \be
 \begin{split}
\la{intuZ} \CZ(L ,R) & =\sum_{m=0}^\infty {(-1)^m\over m!}  
\sum_{r_1, \dots, r_m} {(-1)^{r_1+\dots+ r_m}\over r_1\dots r_m} \int
{du_1\over 2\pi } \dots {du_m\over 2\pi } \\
 &\times \tilde G(u_1^{r_1} ,\dots \, ,u_m^{r_m}) \ e^{- r_1
 \tilde E(u_1)}\dots e^{-r_m \tilde E(u_m)} .
 \end{split}
 \ee
 The change of variables brings a volume-dependent Jacobian (the
 Gaudin determinant)
  \be \la{Gaudindet} \tilde G = \det_{m\times m} \tilde G_{kj} ,
  \qquad \tilde G_{kj} ={\p\over \p u_k} { \tphi _j (u_1^{r_1} ,\dots
  \, ,u_m^{r_m})}, \ee
which gives the density of the particle states in the rapidity space.
The explicit form of the Gaudin matrix $\tilde G_{jk}$ is
\be
\la{Gaudinmatr}
\begin{split}
\tilde G_{kj}& = \(L \tilde p'(u_j) + \sum_{l=1}^m r_l K(u_j, u_l)\)
\d_{jk} - r_kK(u_k, u_j) ,
\end{split}
\ee
where $K(u,v) = {1\over i} \p_u \log \tilde S(u,v)
.
$

\subsection{Graph expansion of the Gaudin determinant}

Let us denote for brevity
\be \tilde p'_j\equiv \tilde p'(u_j) \qquad K_{jk}\equiv K(u_j, u_k).
\ee
 Inspecting the expansion of the Gaudin determinant for $m=1,2,3$
 \be
 \begin{split}
  \tilde G(u^r)& = L \tilde p' \, , \\
\tilde G(u_1^{r_1},u_2^{r_2})&= L^2 \tilde p'_1 \tilde p'_2 +L \tilde
p'_1 r_1 K_{2 1} +L \tilde p'_2 r_2 K_{1 2}, \\
  \tilde G(u_1^{r_1} ; u_2^{_2}, u_3^{ r_3})&= L^3 \tilde p'_1 \tilde
  p'_2 \tilde p'_3\\
& +L^2 \tilde p'_2 \tilde p'_3 r_2 K_{1 2}+L^2 \tilde p'_2 \tilde p'_3
r_3 K_{1 3} +L^2 \tilde p'_1 \tilde p'_3 r_1 K_{2 1} \\
 & +L^2 \tilde p'_1 \tilde p'_3 r_3 K_{2 3}+L^2 \tilde p'_1 \tilde
 p'_2 r_1 K_{3 1}+L^2 \tilde p'_1 \tilde p'_2 r_2 K_{3 2} \\
 &+\tilde p'_3 L r_1 r_3 K_{1 3} K_{2 1}+\tilde p'_3 L r_2 r_3 K_{1 2}
 K_{2 3}+\tilde p'_3 L r_3^2 K_{1 3} K_{2 3}\\
 &+\tilde p'_2 L r_1 r_2 K_{1 2} K_{3 1}+\tilde p'_1 L r_1^2 K_{2 1}
 K_{3 1}+\tilde p'_1 L r_3 r_1 K_{2 3} K_{3 1} \\
 &+\tilde p'_1 L r_2 r_1 K_{2 1} K_{3 2}+\tilde p'_2 L r_2^2 K_{1 2}
 K_{3 2}+\tilde p'_2 L r_2 r_3 K_{1 3} K_{3 2},
 \end{split}
 \ee
we see that there are no cycles of the type $K_{12}K_{21}$ or
$K_{12}K_{23}K_{31}$.  We will see below that this property hold for
general order $m$.  To evaluate the Gaudin determinant for general
state $\{u_1^{r_1}, \dots, u_m^{r_m} \}$, we will consider in the
following a slightly modified Gaudin matrix, $\hat G_{kj}= \tilde
G_{kj} r_j$.  The determinants of the two matrices are simply related,
\be \tilde G= {\det\hat G_{jk} \over \prod_{j=1}^m r_j}, \qquad \hat
G_{kj} \equiv \tilde G_{kj} r_j.  \ee
The the modified Gaudin matrix has the advantage that it is a sum of a
diagonal matrix $\hat D_{j}\d_{jk}$ and a Laplacian matrix $ \hat
K_{kj}$ (a matrix with zero row sums):
\be\begin{split}
\la{decompoGau}
&\hat G_{kj}=  \hat D_k \, \d_{kj} - \hat K_{kj}
\\
& \text{with}\ \ 
\hat D_j = L r_j \tilde p'(u_j) \ \ \text{and}\ \  
  \hat K_{k,j} = r_k  r_j  K(u_k, u_j) - \d_{kj} \sum _{l=1}^m  r_j  r_l  K(u_j,u_l)
\end{split}
\ee
 According to the {\it Matrix-Tree Theorem} (see e.g.
 \cite{Matrix-Tree , ABDESSELAM200451}), the determinant of the
 matrix $\hat G_{ij}$ can be expanded as a sum of graphs called {\it
 directed spanning forests}.  A directed forest spanning the graph
 $\G$ is an oriented subgraph $\CF$ fulfilling the following three
 conditions:
      
 \noindent 
 (i)\ \ \  $ \CF $ contains all vertices  
 of $\G$;
 
   \noindent 
  (ii)  \ \  $\CF$ does not contain cycles; 

  \noindent (iii) \ For any vertex of $\G$ there is at most one
  oriented edge of $\CF$ ending at this vertex.

\begin{figure}[h!]
         \centering
         \begin{minipage}[t]{0.8\linewidth}
            \centering
            \includegraphics[width=11cm]{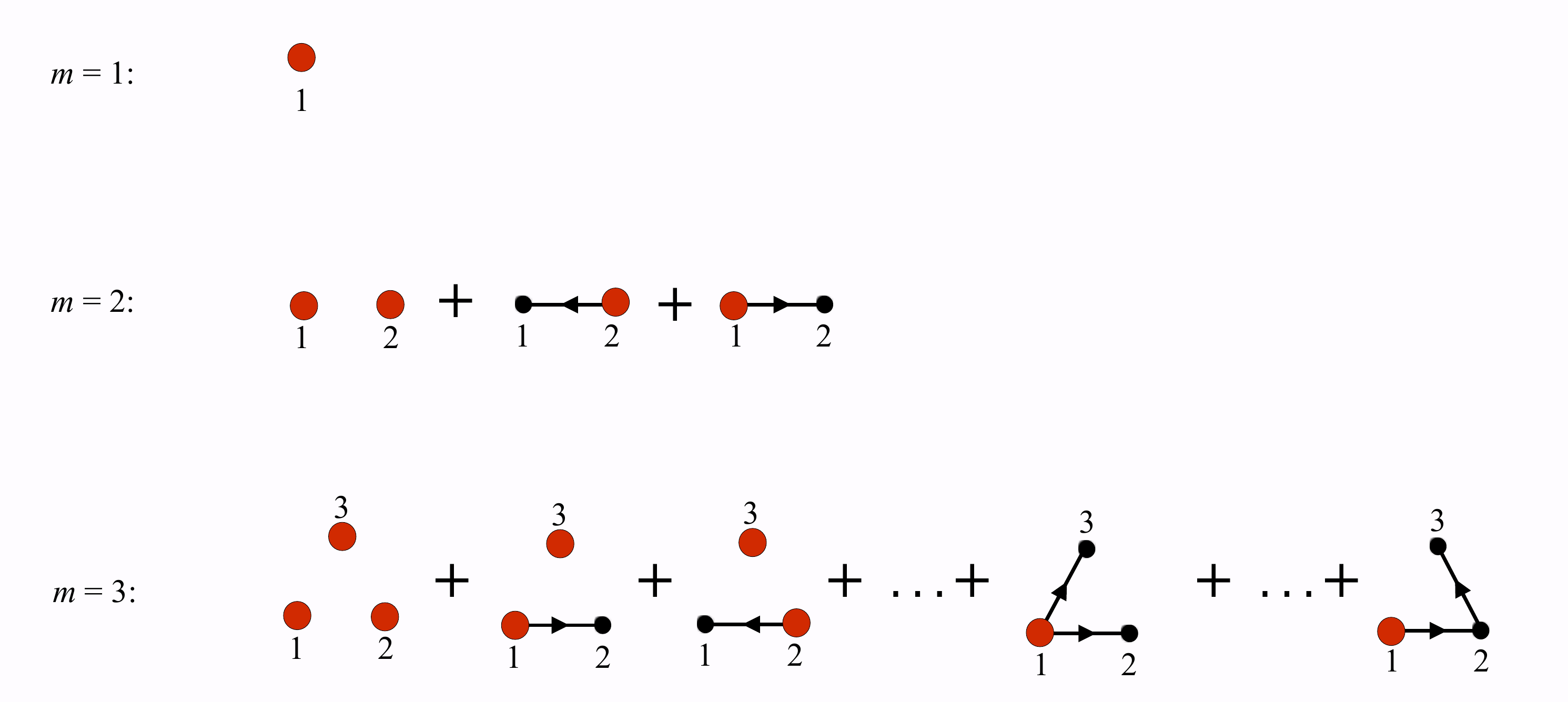}
	  \caption{ \small   
	The expansion of the  determinant of the matrix 
 defined in eq.  \re{decompoGau}  in directed spanning forests
  for $m=1,2,3$.  Ellipses mean sum over the permutations of the
  vertices of the preceding graph.
  Each vertex of a directed tree, except for the root,
 has exactly one incoming edge  and an arbitrary number of 
 outgoing edges. The root  can have only outgoing  edges.
 A factor $\hat K_{kj} $ is associated with each edge $\ell_{jk}$.
A factor $\hat D_k$ is associated with the roots of each connected tree, which is
symbolised by a red dot.  }
\label{fig:Matrix-tree}
         \end{minipage}
           \end{figure}

The vertices with no incoming lines are called {\it roots}.  Any
forest $\CF$ is decomposed into connected components called {\it
directed trees}.  Each tree contains one and only one root.  The
Matrix-Tree Theorem states that the determinant of the matrix $\hat G$
is a sum of all directed forests $\CF$ spanning the totally connected
graph with vertices labeled by $j=1,\dots, m$:
\be \la{MTexpansioin} \det _{m\times m} \( \hat D_j \d_{jk} - \hat
K_{jk}\)= \sum_{ \CF } \prod_{v_i\in \text{roots }} \hat D_i \
\prod_{\ell_{jk}\in\CF} \hat K_{kj}.  \ee
 The weight of a forest $\CF$ is a product of factors $\hat D_k$
 associated with the roots and factors $ \hat K_{kj}$ associated with
 the oriented edges $\ell_{jk}= \< v_j\to v_k\>$ of the $\CF$.  The
 expansion in spanning forests for $m=1,2,3$ is depicted in Fig.
 \ref{fig:Matrix-tree}.

\begin{figure}[h!]
         \centering
         \begin{minipage}[t]{0.7\linewidth}
            \centering
            \includegraphics[width=8.9cm]{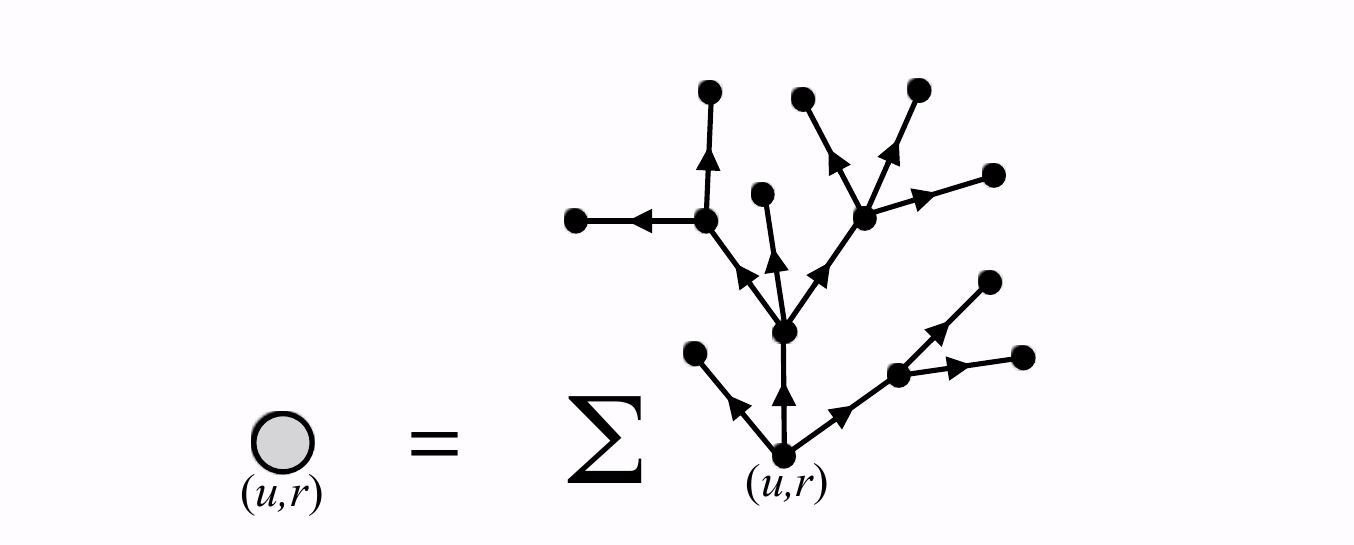}
	  \caption{ \small  The generating function $\tilde Y_r(u)$ of the directed
trees with root at $(u, r)$.  The weight of each tree in the sum is a
product of factors associated with its vertices and edges according to
the Feynman rules \re{Feynmp}.  
The root is denoted by a black dot because here it has the same weight as the 
rest of the vertices of the tree.}
\label{fig:Yr}
         \end{minipage}
           \end{figure}

 Applying the above graph expansion to the Jacobian, we write the
 partition function as
 \be
 \begin{split}
   \la{intuZb} \CZ(L ,R) & =\sum_{m=0}^\infty {(-1)^m\over m!}
   \sum_{r_1, \dots, r_m}\ \int \prod_{j=1}^m {du_j\over 2\pi} \
   {[-e^{- R\tilde E(u_j)}]^{r_j }\over r_j ^2} \\
 &\times \sum_{ \CF } \prod_{j\in\text{roots}} Lr_j \tilde p'(u_j)
 \prod_{ \ell_{ij}\in\CF} r_i r_jK(u_j,u_i) .  \end{split} \ee
The next step is to invert the order of the sum over graphs and the
integral/sum over the coordinates $(u_j, r_j)$ assigned to the
vertices.  As a result we obtain a sum over the ensemble of abstract
oriented tree graphs, with their symmetry factors, embedded in the
space $\IR\times \IN$ where the coordinates $u,r$ of the vertices take
values.  The embedding is free, in the sense that the sum over the
positions of the vertices is taken without restriction.  As a result,
the sum over the embedded tree graphs is the exponential of the sum
over connected ones.  One can think of these graphs as tree level
Feynman diagrams obtained by applying the following Feynman rules:
  \medskip \medskip 
 \be
 \boxed{\begin{split} \la{Feynmp} 
 \\
 \imineq{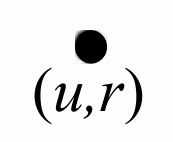}{5} \  \qquad  &= \  \  \ {(-1)^{r-1} \over r^2} \ e^{-r R\tilde E(u)}
\\
\\
 \imineq{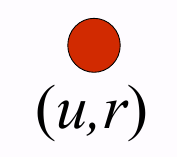}{5} \ \ \qquad  &= \  \  \  L p'(u)\, {(-1)^{r-1} \over r} \ e^{-r R\tilde E(u)}
\\
\\
 \imineq{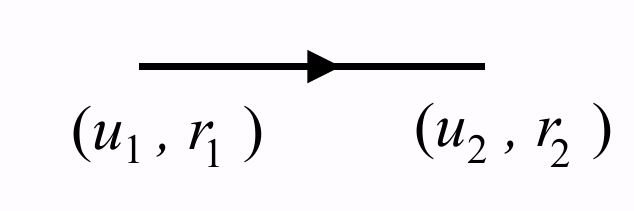}{8} \hskip-3mm & \  \ \ = \ \ \ r_1 r_2 K(u_2, u_1)
 \end{split} 
 }\ee
 In this way we can write the free energy as
 \be
 \begin{split}
\la{intuZbFE} \log \CZ(L ,R) &   = L \int {du\over 2\pi} \tilde
p'(u) \sum_{r=1}^\infty r \tilde Y_r(u),
 \end{split}
 \ee
where $\tilde Y_r(u)$ is the partition sum of all {\it connected}
directed rooted trees with root at the point $(u, r)$, fig.
\ref{fig:Yr}.

Eq.  \re{intuZbFE} gives the free energy up to $e^{-L}$ terms, hence
the subleading terms in the expansion \re{FREEXP} vanish.  Of course
this is true only for periodic boundary conditions.
Eq.  \re{intuZbFE} gives the free energy up to $e^{-L}$ terms, hence
the subleading terms in the expansion \re{FREEXP} vanish.  Of course
this is true only for periodic boundary conditions.
 \begin{figure}[h!]
         \centering
         \begin{minipage}[t]{0.7\linewidth}
            \centering
\includegraphics[scale=0.7]{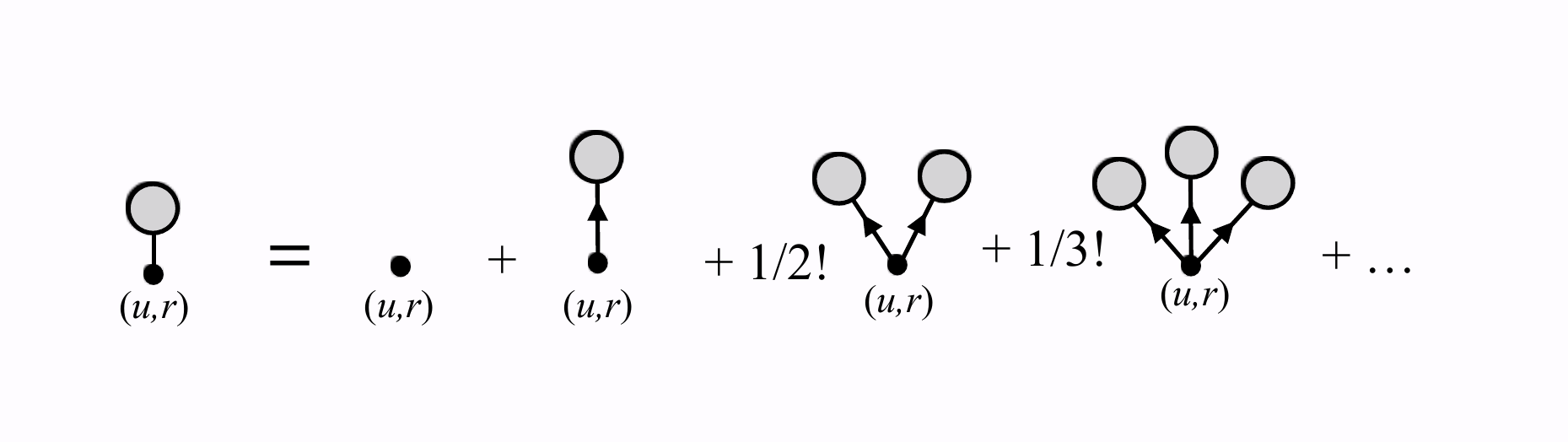}
\vskip -1cm
 \caption{\small\ The non-linear equation for the generating function
 $\tilde Y_r(u)$ of the trees with root at $(u,r)$ }
\label{fig:eqYr}
    \end{minipage}
           \end{figure}%

\subsection{Performing the sum over trees}

As any partition sum of trees, $\tilde Y_r(u)$ satisfies a simple
non-linear equation (a Schwinger-Dyson equation in the QFT language)
depicted in Fig.  \ref{fig:eqYr},
\be
\la{nonlineq}
\begin{split}
\tilde Y_r(u) &={(-1)^{r-1}\over r^2} e^{- r R\tilde
E(u)}\sum_{n=1}^\infty {1\over n!} \( \sum_{s }\int {dv \over 2\pi} rs
K(v , u) \tilde Y_{s }(v )\)^n \\
&={(-1)^{r-1}\over r^2} \[ e^{- R\tilde E(u)}\ e^{\sum_{s} \int {dv
\over 2\pi} s K(v , u) \tilde Y_{s }(v )}\]^r.  \end{split} \ee
 In particular  for $r=1$ 
 \be \la{Yone} \tilde Y_1(u) = e^{- R\tilde E(u)}\ e^{ \sum_s\int {dv
 \over 2\pi} s K(v , u) \tilde Y_{s }(v )}.  \ee Substituting the rhs
 of \re{Yone} in the square brackets in the second line of eq.
 \re{nonlineq}, we express all $Y_r$ in terms of $Y_1$,
\be\begin{split} \tilde Y_r(u) &={(-1)^{r-1} \over r^2} [\tilde
Y_1(u)]^r, \quad r=1,2,3, \dots .
 \end{split} 
 \la{YrY1}\ee
Now we can express the rhs of \re{intuZbFE} and the exponent in on the
rhs of \re{Yone} in terms of $Y_1$ only,
\be \la{vacuumtrees} \sum_r r\tilde Y_r(v)= \log\[ 1+ \tilde Y_1(v)\].
\ee
 Now eq.  \re{Yone} becomes a closed equation for $Y_1$,
\be
\begin{split}
\la{Ysyscentral} \tilde Y_1(u) &= e^{-R \tilde E(u) +\int { dv\over
2\pi} K(v,u) \log\[1+ \tilde Y_1(v)\]},
\end{split}
\ee
 which determines completely the free energy
 \be
 \begin{split}
\la{intuZbFEb}
\log \CZ(L ,R)
& =  L
 \int {du\over 2\pi}
 \tilde p'(u) \log\[ 1+ \tilde  Y_1(v)\] 
 + o(e^{-L}).
 \end{split}
 \ee
In this way we reproduced, by summing up the tree expansion of the
free energy, the TBA equation for the pseudoenergy $\e(u) = - {1\over
L} \log \tilde Y_1(u)$.  The expression \re{intuZbFEb} for the free
energy is true in all orders in $1/L$.  In particular, there is no
$O(1)$ piece, in accord with the TBA based computation in
\cite{Pozsgay:2010tv}.

  \section{   The  energy of an excited state}
  \label{section:Casimir}

 In this section we will apply the tree expansion to the case of an
 excited state $|\uu\>$ in the physical channel characterised by a set
 of rapidities $\uu= \{u_1, \dots, u_N\}$.  We assume that the excited
 state is an eigenstate of the Hamiltonian with energy given by eq.
 \re{physenergy}.
 
  For large $R$ the wrapping phenomena can be neglected and the
  rapidities $\uu$ satisfy the asymptotic Bethe equations
  \re{ABAb}-\re{Betheq}.
 In order to determine the exact energy and the exact values of the
 rapidities for finite $R$, we again introduce a cutoff $L$ by
 compactifying the cylinder into a torus obtained as the product of a
 space-like circle $R$-circle and a time-like $L$-circle, with a
 projector $|\uu\>\<\uu|$ inserted in the physical channel.  The
 phases of the mirror particles now contain an extra piece which comes
 from the scattering with the physical particles:
\be
\label{Betheqb}
  \tilde\phi_j (v_1, \dots , v_M)\equiv \tp(v_j) L + {1\over i}\sum_{k
  =1}^ N \log S(\tv_j,u_k) + {1\over i}\sum_{l(\ne j)}^ M \log
  S(\tv_j,\tv_l), \qquad j=1,\dots, M. \ee

 The computation of the partition function then follows strictly the
 argument of the previous section, with the only difference that the
 mirror energy is modified by the scattering with the physical
 particles.  We have to replace
 \be e^{-L \tilde E(v)} \ \to \tilde Y_1^\circ(v) \equiv e^{-L \tE(v)}\,
 \prod_{k=1}^M S(\tv, u_k) \,.  \ee
 Furthermore we have to add to the free energy the 
 contribution from the physical particles
 that go directly to the opposite edge without scattering,
 \be
 \begin{split}
\la{intuZbFExst} \log \CZ(L,R, \uu) & = - L \sum_{j=1}^N E(u_j) + L
\int {du\over 2\pi}
 \tilde p'(u) \log\[ 1+ \tilde Y_1(v)\] 
 + O(e^{-L}).
 \end{split}
 \ee
 with the function $Y(u)$ satisfying non-linear integral equation
 which slightly generalises eq.  \re{Ysyscentral},
 \be \la{nlfebis} \tilde Y_1 (v) &= \tilde Y_1^\circ(v)\, e^{ \int
 {du\over 2\pi} \log(1+ \tilde Y _1(u) ) K (u,v)} .  \ee

\begin{figure}[h!]
         \centering
         \begin{minipage}[t]{0.8\linewidth}
            \centering
            \includegraphics[width=8.5cm]{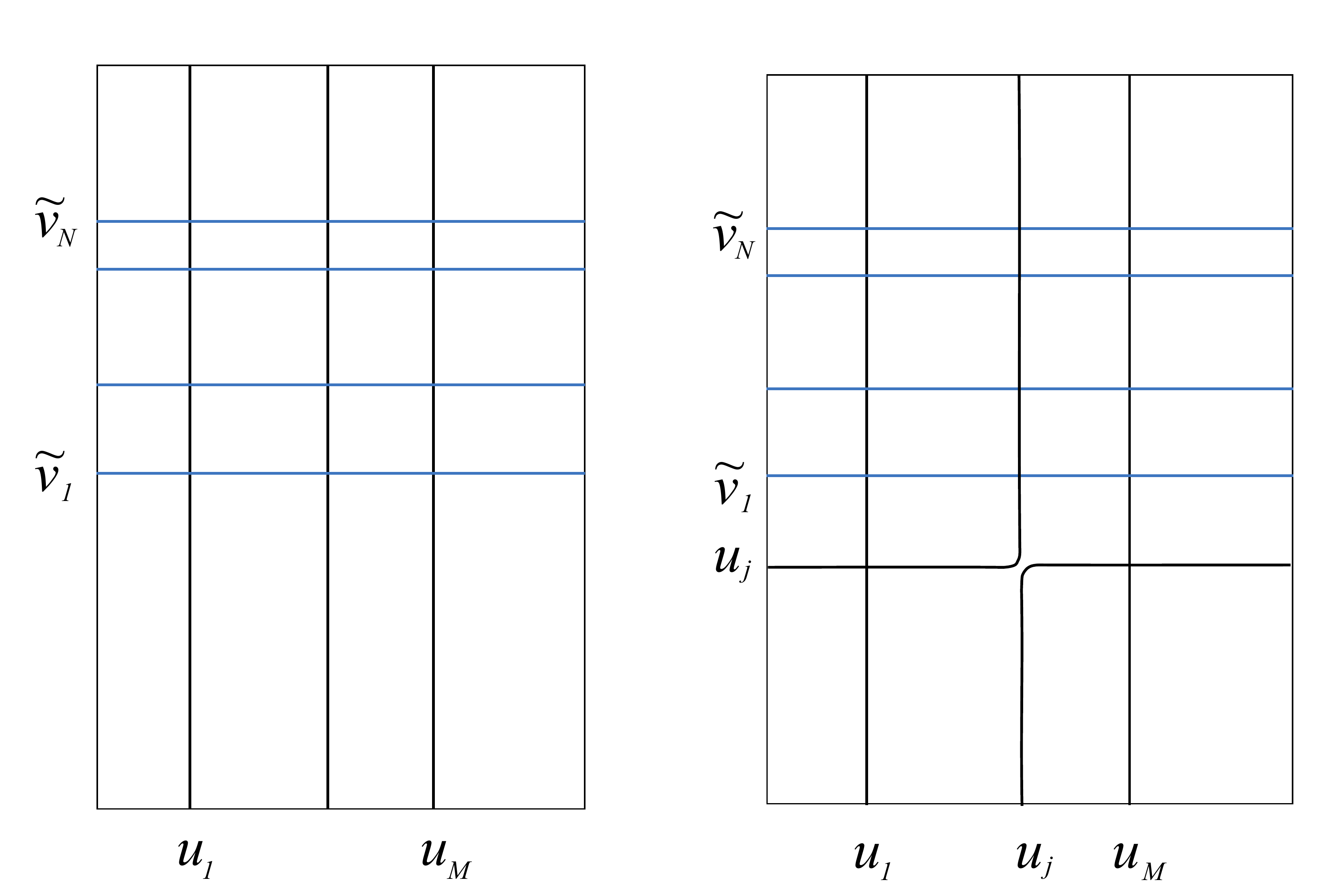}
            \centerline{a\qquad       \qquad \qquad \qquad \qquad \quad b}
	  \caption{ \small  The configurations that lead to the exact Bethe
equation.  The physical magnon winding once around the space circle
has the same effect, up to a factor $(-1)$, as a physical magnon going
straight in presence of a mirror magnon with rapidity $u_j$.}
\label{fig:BAExact}
         \end{minipage}
           \end{figure}

The rapidities of the physical particles are no longer determined by
the asymptotic Bethe-Yang equations but by the ``exact Bethe
equations'' which take into account all virtual excitations in the
mirror channel.  The exact Bethe equations are formulated in terms of
the function $\tilde Y_1$.  In order to avoid confusion we introduce
the Y-function in the physical channel, which is related to $\tilde Y$
by
  \be \tilde Y_1(v) = Y_1(\tilde v).  \ee
The exact Bethe equations are obtained by the following requirement.
Let $\CZ_j (R,L)$ be the partition function with the $j$-th physical
particle winding once around the space circle before winding around
the time circle.  The configurations that contribute to $\CZ(R,L)$ and
$\CZ_j (R,L)$ are depicted in Figs.  \ref{fig:BAExact}a and
\ref{fig:BAExact}b.

In order to compute the partition function $\CZ_j (R,L)$ we notice
that the configurations in Fig.  \ref{fig:BAExact}b can be simulated
by pulling one of the mirror particles out of the thermal ensemble
giving to its rapidity a physical value $ u_j$.  Indeed, since
$S(u_j,u_j)=-1$, the partition function in presence of such extra
mirror particle is $-\CZ_j (R,L)$.  In this way $\CZ_j (R,L)$ is given
by the sum over all trees, with one extra tree having a root $\tilde v
= u_j$ and $r=1$.  The generating function for such trees is $Y_1(u_j
)$, while the contribution of the ``vacuum'' trees give the partition
function: $\CZ_j = - Y_1(u_j)\, \CZ$.  The periodicity in the space
direction requires that $\CZ_j=\CZ$, which gives the exact Bethe-Yang
equations
 \be
 Y_1(u_j)=-1, \quad j=1, \dots, N.
 \ee

\section{One-point functions at finite volume/temperature}
\label{section:LM}

In this section we will apply the tree expansion to compute the
diagonal matrix elements of a local operator at finite volume $R$.
The LeClair-Mussardo conjecture \cite{Leclair:1999ys} gives an
expression for the exact finite temperature one-point functions.  In
terms of infinite-volume diagonal connected form factors, and
densities of mirror states determined by the TBA equation.  The
conjecture was proven for operators representing densities of
conserved quantities in \cite{Saleur:1999hq} and for general local
operator in \cite{Bajnok:2017bfg}.  The proof of \cite{Bajnok:2017bfg}
concerns the formula about the diagonal form factors in asymptotically
large volume conjectured by Pozsgay and Takacs \cite{Pozsgay:2007gx},
which is equivalent to the L-M formula.   The Pozsgay-Takacs formula,
which generalises a result by Saleur \cite{Saleur:1999hq},
gives an expansion of the diagonal matrix elements of a local operator
in terms of the infinite-volume form factors with the same or lower
number of particles.

 \subsection{The one-point function in terms of connected diagonal form factors}

 In order to simplify the notations, in this section we assume that
 the physical Hilbert space is associated with the $L$-circle and the
 mirror Hilbert space is associated with the $R$-circle.  In infinite
 volume, all matrix elements of a local operator $\CO$ can be
 expressed, with the help of the crossing formula, in terms of the
 {\it elementary form factors}
   \be \la{elemFF} F_n^\CO(u_1, \dots, u_n) = \< 0| \CO|u_1, \dots,
   u_n\>_{\infty}.  \ee
The elementary form factors for local operators satisfy the Watson
equations
\be F_n(u_1, \dots, u_j, u_{j+1}, \dots, u_n)=S(u_j, u_{j+1}) F_n(u_1,
\dots, u_{j+1}, u_{j}, \dots, u_n) \ee and have kinematical
singularities
\be \la{kinsing} F( v, u, u_1, \dots, u_n) = {i\over \bar v-u} \( 1-
\prod_{j=1}^n S(u, u_j) \) F_n(u_1, \dots, u_n) + \text{regular}, \ee
where $\bar v$ is obtained from $v$ by a crossing transformation.
Here it is assumed that the infinite volume states are normalised as
$\<u|v\> = 2\pi \d(u-v)$.

The diagonal limit of the form factors for local operators is
ambiguous\footnote{In the case of the non-local operators the
situation is even worse: their diagonal limit diverges as $L^M$ where
$M$ is the number of the particle pairs.} and there are two
prescriptions for evaluating the finite piece, the symmetric and the
connected one \cite{Pozsgay:2007gx}.  \def\ve{{\varepsilon}} The
connected diagonal form factor $F_{2n}^c(u_1, \dots, u_n) $ is
obtained by performing the simultaneous limit $\ve_1, \dots, \ve_n\to
0$ of the elementary form factor $F_{2n}(u_1, \dots, u_{2n})$ defined
by eq.  \re{elemFF}, with $u_{2n-j+1} = \bar u_j + i\ve_j$.  The limit
is not uniform and depends on the prescription, which in this case is
to retain only the $\ve$-independent part:
\be F_{2n}^c(\bar u_n +i\ve_n, \dots, \bar u_1+ i\ve_1, u_1, \dots,
u_n) = F_{2n}^c(u_1, \dots, u_n) + \ \ve\text{-dependent terms}.  \ee

The   Saleur-Pozsgay-Takacs formula   \cite{Saleur:1999hq, Pozsgay:2007gx}  
 relates the diagonal
matrix elements in asymptotically large but finite volume $L$ to the
connected diagonal form-factors.  The formula reads
%
%
\be \la{PTexpansion} \< \uu |\CO | \uu\> _{_L} =
{
\sum_{\a\cup\bar\a = \uu} F^c_{2|\a|}(\a) \times
\det_{j,k\in \bar\a} G_{jk} 
}+ O(e^{-L}), 
\ee
where the sum goes over all partitions of the rapidities $\uu=\{ u_1, \dots,
u_n \}$ in to two complementary sets $\a$ and $\bar \a$, and $G_{jk} =
\p_{u_j}\phi_k$ is the Gaudin matrix for the $n$ rapidities.  
It is assumed that $F^c_0=0$, so there is no term with $\a=\emptyset$. 
The  formula   is written for the normalisation  with the Gaudin norm
 \be
 \<\uu|\uu\> = \det_{j,k\in \uu} G_{jk}.
 \ee
 The
determinants on the rhs are the minors of the Gaudin determinant
obtain by deleting the lines and the columns that belong to the subset
$\a$.  It is shown \cite{Pozsgay:2010xd, Pozsgay:2010tv} that the
expansion \re{PTexpansion} is equivalent to the Leclair and Mussardo
series for the one-point function of a local operator
\cite{Leclair:1999ys}
 \be \la{LMseries} \< \CO \>_R = \sum_{n=1}^\infty {1\over n!} \int
 \prod_{j=1}^n {du_j\over 2\pi } f(u_j) \ F^c_{2n}(u_1, \dots, u_n) ,
 \qquad f(u) = {Y_1(u)\over 1+Y_1(u)}.  \ee

Below we will derive the Leclair-Mussardo formula from the tree
expansion method.  In particular, we will reproduce the result
obtained by Saleur \cite{Saleur:1999hq} for the one-point function of
a conserved charge.  For that we will need the diagonal matrix
elements also for the multi-wrapping states $|u_1^{r_1}, \dots , u_m
^{r_m}\>$.  We will make a very natural conjecture about this action,
which turns out to be compatible with the correct formula
\re{LMseries}, namely
\be \la{PTexpansionwrap} \< u_M^{r_M},\dots, u_1^{r_1}|\CO |
u_1^{r_1},\dots, u_M^{r_M}\> _L = {\sum_{\a\cup\bar\a = \{ u_1, \dots,
u_M\}} \prod_{j\in\a} r_j \, F^c_{2|\a|}(\a) \times \det_{j,k\in
\bar\a} G_{jk} }
.
\ee
The logic behind this conjecture is that the action of the operator on
a multi wrapping particle is the same as if it were single wrapping
particle.  The only difference is that the $r$-wrapping particle
appears $r$ times in the same time slice, the operator acts on each
copy, which brings an overall factor of $r$.  We should mention here
that a discussion about the ``multi-diagonal'' matrix elements was
presented in \cite{Bajnok:2017mdf}.
     
 \subsection{LeClair-Mussardo series  from the tree expansion}

Repeating the argument from the beginning of section
\ref{subsection:mntorap}, we can perform the sum over the complete set
of states in the thermal expectation value of the operator $\CO$
  \be \la{thexpvO} \< \CO \>_R = \sum_{M=0}^\infty \ \
  \sum_{n_1<n_2<\dots <n_M} e^{- R E(n_1,\dots, n_M)} \< n_1, \dots,
  n_M|\CO|n_M, \dots, n_1\> \ee
by inserting the expansion \re{PTexpansion} in each term of the sum
and proceeding as in Section \ref{subsection:modenumbers}.  The
expansion analogous to the formula \re{intuZ} for the partition
function is
\be\begin{split} \la{ThermalO} \< \CO\>_{ R} &= {1\over \CZ(L,
R)}\sum_{m=0}^\infty {(-1)^m\over m!}\!\!\!  \sum_{r_1, \dots, r_m}\
\int {du_1\over 2\pi}\dots {du_m\over 2\pi} \ \, {e^{- Lr_1 E(u_1) - L
r_m E(u_m)}\over r_1\dots r_m} \\
  & \times \sum_{\a\cup\bar\a = \{ u_1, \dots, u_m\}} \prod_{j\in \a}
  r_j \ F^c_{2|\a|}(\a) \ {\det_{j,k\in \bar\a} \hat G_{jk} \over
  \prod_{i\in \bar\a }r_i},
\end{split}
\ee
where the matrix $\hat G_{jk}$ is defined by eq.  \re{decompoGau} with
$\tilde p $ replaced by $p$ and the scattering kernel defined as
$K(u,v)= {1\over i}\p_u \log S(u,v)$.

\begin{figure}[h!]
         \centering
         \begin{minipage}[t]{0.8\linewidth}
            \centering
            \includegraphics[width=9.9cm]{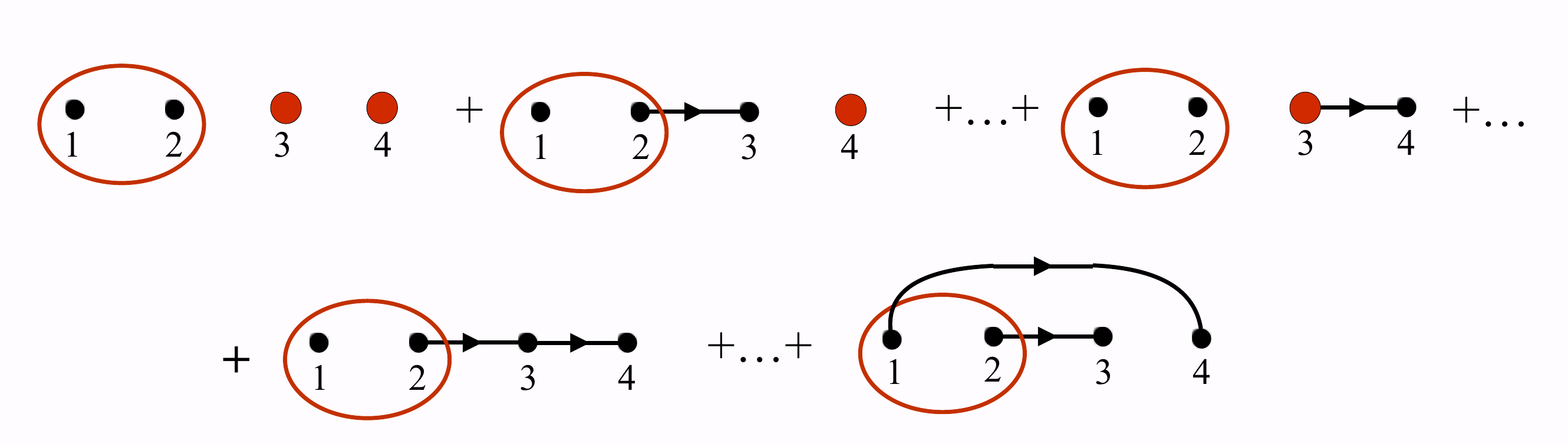}
	  \caption{ \small The tree expansion for a principal minor of the
 Gaudin matrix  $\det _{ j,k\in\bar\a} \hat G_{jk}$  for  $\a=\{ 1,2\} $ and $ \bar\a = \{ 3,4\}$.  }
\label{fig:Minors}
         \end{minipage}
           \end{figure}

 The next step is to apply the matrix-tree theorem for the diagonal
 minors of the Gaudin determinant in the last factor in the integrand
 in \re{ThermalO}.  A minor obtained by removing all edges and all
 columns from the subset $ \a\subset\{1, \dots, m\}$ of the matrix
 $\hat G_{jk}$ defined in eq.  \re{decompoGau} has the following
 expansion,
\be \la{MTexpansioinM} \det _{ j,k\in\bar\a} \hat G_{jk}= \sum_{ \CF
\in \CF_{\a, \bar\a}} \prod_{\text{roots}\in \bar\a} \hat D_i \
\prod_{\ell_{jk}\in\CF} \hat K_{kj}.  \ee
The spanning forests $\CF\in \CF_{\a, \bar\a} $ are subjected to
conditions $(i)-(iii)$ of section 2.5, with the additional restriction
that all vertices belonging to $ \a$ are roots.  The weight of these
roots is one.  An example is given in fig.  \ref{fig:Minors}.

The
expansion \re{MTexpansioinM} follows directly from the expansion
\re{MTexpansioin} of the previous section which corresponds to the
particular case $\a=\emptyset, \bar\a = \{ u_1, \dots, u_m\}$.
Indeed, the rhs of \re{MTexpansioinM} by retaining only the terms in
the rhs of \re{MTexpansioin} that contain the factor $\prod_{j\in \a}
\hat D_j$ and then dividing the sum by this factor.

Now we can proceed similarly to what we have done in the computation
of the partition function, where rearranging of the order of summation
allowed us to rewrite the sum as a series of tree Feynman diagrams.
This time there will be two kinds of Feynman graphs: the ``vacuum
trees'' and diagrams representing a vertex $F^c_{2n}$ with $n$ lines
and a tree attached to each line.  The weight of such tree is the same
as the weight of the vacuum trees except for a factor of $r^2$
associated with the root.  This factor becomes obvious if one writes
the dependence of the integrand/summand of \re{ThermalO} on the
wrapping numbers $r_1, \dots, r_m$ as
 $$
 {1\over r_1^2\dots r_m^2} \prod_{j\in\a} r_j^2.
 $$
 The sum over the vacuum trees cancels with the partition function and
 the sum over the surviving terms has the same structure as
 \re{LMseries}, which is depicted in Fig.  \re{fig:LMseries}.  The
 factor $f(u)$ is obtained as the sum of all trees with a root at the
 point $u$, with extra weight $r^2$ associated with the root:
 \be\begin{split} \la{seriesforf} \sum_r r^2 Y_r(u)= \sum_r
 {(-1)^{r-1} } [ Y_1(u)]^r ={ Y_1(u) \over 1+ Y_1(v)}= f(u) .
 \end{split} 
 \la{fuuu}
 \ee
The difference of the sum over trees in the factor $f(u)$ compared
with the sum over vacuum trees \re{vacuumtrees} is that there is an
extra factor $r$ associated with the root reflecting the breaking of
the $Z_r$ symmetry of the corresponding wrapping process.

\begin{figure}[h!]
\begin{center}
\includegraphics[scale=0.59]{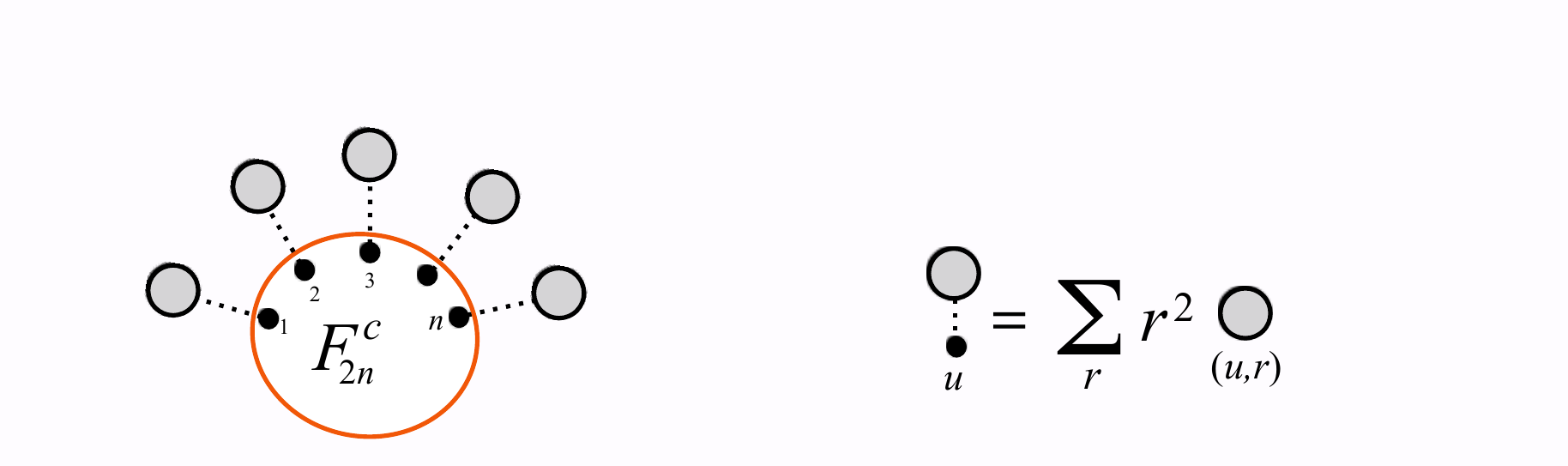}
 \caption{\small\ The tree expansion for the thermal expectation value
 \re{PTexpansionwrap}  
 of a local operator.}
\label{fig:LMseries}
\end{center}
\end{figure}

   \subsection{The case of a conserved charge}

 The simplest local operator $\CO$ is of the type of conserved charge,
 such as the energy or the momentum.  Such operators act diagonally on
 multi-particle states with one-particle values $ o (u)$.  The matrix
 elements of the operator on a multi-particle state at zero
 temperature are
\be \la{defO} \CO= L^{-1}\int dx \CO(x),\qquad { \< u_n,\dots,
u_1|\CO|u_1, \dots, u_n\> \over \< u_n,\dots, u_1|u_1, \dots, u_n\> }
= {1 \over L} \sum_{j=1}^n o(u_j).  \ee
By direct computation one obtains \cite{Saleur:1999hq}
\be\begin{split} F^c_{2n}(u_1, \dots, u_n) &= p'(u_1) K(u_2,
u_1)K(u_4, u_3)\dots K(u_n, u_{n-1}) \ o(u_n) \\
&+\mathrm{permutations},
\end{split}
\ee
to be substituted in the LeClair-Mussardo series \re{LMseries}.

\begin{figure}
         \centering
         \begin{minipage}[t]{0.8\linewidth}
            \centering
            \includegraphics[width=11.9cm]{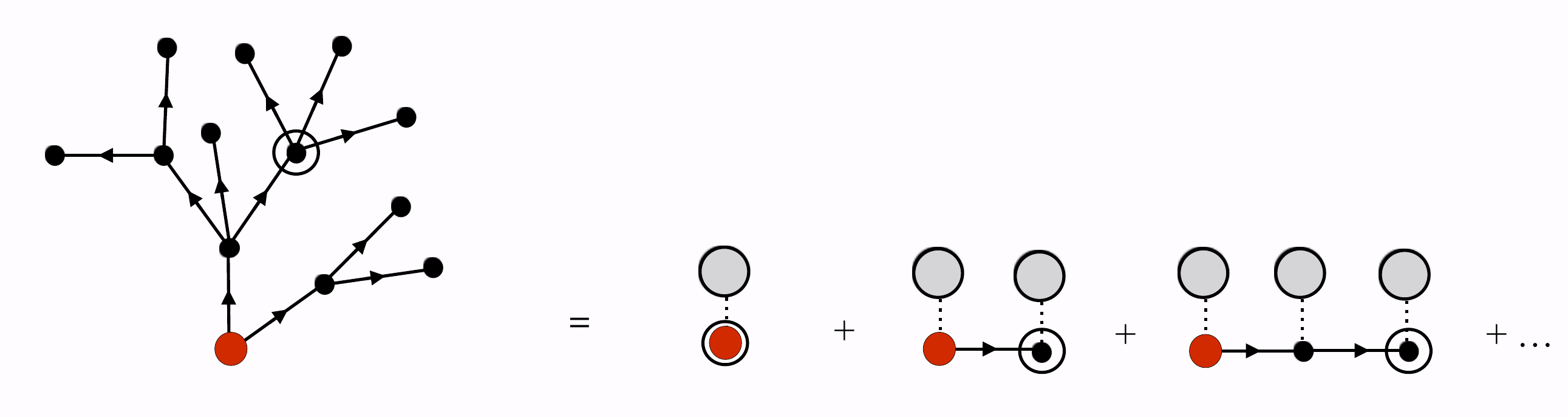}
	  \caption{ \small  
	  The factor $ Y(u_1, r_1; u_2, r_2) $
 in the  tree expansion for the thermal expectation value
 of a conserved charge.   
 The circle  symbolises the vertex where the one-particle 
 operator $o(u)$ is inserted.
	   }
\label{fig:CCharge}
         \end{minipage}
           \end{figure}

This formula can be readily obtained from the tree expansion using
only the definition \re{defO}.  We start with the series for the
partition function \re{intuZb}, with $\tilde p$ and $ \tilde E $
replaced by $p$ and $E$, and multiply each term by the eigenvalue of
the operator $\CO$, which acts on the states $| u_1^{r_1} \dots,
u_m^{r_m}\>$ as
 \be \CO| u_1^{r_1} \dots, u_m^{r_m}\> = {1\over L} \sum_j r_j o(u_j)\
 | u_1^{r_1} \dots, u_m^{r_m}\>.  \ee
 After expanding the Gaudin norm in trees, one of the trees will
 acquire an extra factor $r_j o(u_j)$ associated with one of its
 vertices.  The sum over the vacuum gives the partition function which
 is to be stripped off and one is left with the sum over connected
 trees with one marked point,
 \be
 \begin{split}
\la{intuZOTb} \< \CO\>_{L, R} &
= \int {du_1\over 2\pi} \int {du_2\over
2\pi} \sum_{r_1, r_2 } L r _1p'(u_1) \, Y(u_1, r_1; u_2, r_2) {1\over
L}r_2o (u_2)
 \end{split}
 \ee
where $ Y(u_1, r_1; u_2, r_2) = \d(u_1-u_2) \d_{r_1, r_2}Y_{r_1}(u_1)
+\dots $ is the partition function of all
directed trees with root at $(u_1,r_1)$ and a marked vertex at $(u_2,
r_2)$.  Any such tree can be decomposed into a backbone consisting of
the edges connecting the root and the marked point, and a collection
of trees rooted at the vertices along the backbone.  We will associate
a factor $K_{jk}$ with the edge $\ell_{kj}$ of the backbone, while the
factors $r_k$ and $r_j$ will be absorbed into the weights of the trees
rooted at the vertices $k$ and $j$.  In this way the trees rooted at
the point $j$ of the backbone contain a factor $r_j^2$ coming from the
two adjacent edges.  The sum of such trees gives the factor $f(u) $,
eq.  \re{seriesforf}.  The net result is
 \be \la{LMseriesCC} \< \CO \>_R = \sum_{n=1}^\infty \int \prod_{j=1}^n
 {du_j\over 2\pi } \ p'(u_1) f(u_1) K(u_2, u_1)f(u_2) K(u_3, u_2)
 \dots K(u_{n}, u_{n-1}) f(u_n) o(u_n) \ee
which is illustrated by fig.  \ref{fig:CCharge}

Another way to obtain the one-point function of a conserved charge is
by replacing the energy $E(u)$ in the thermal factors with $ E(u) - \a
o(u)$.  In this way the problem is reduced to the problem of the
computation of the thermal partition function, but with slightly
changed form of the energy.  Since the computation of the partition
function does not depend on the specific form of the energy, we can
use the formulas of the previous section where $Y_1(u)$ is replaced by
$Y_1(u,\a)$ determined by the non-linear integral equation
\be
\begin{split}
\la{YsyscentralO}
  \log Y(u,\a)  
&= - R E(u) + \a o (u) +\int { dv\over 2\pi} K(v,u) \log\[1+
Y_1(v,\a)\].
\end{split}
\ee
The one-point function is given by the derivative
 \begin{align}
\label{Otherm}
 \< \CO\>_{ R} = {\p\over \p\a} \int {du\over 2\pi} p'(u) \log(1+
 Y_1(u,\a)) \Big|_{\a=0} = \int {du\over 2\pi} p'(u) f(u) \tilde o (u)
 ,
\end{align}
with $\tilde o(u)$ satisfying a linear integral equation obtained by
differentiating \re{Ysyscentral},
 \be \tilde o (u) = o (u) + \int { dv \over 2\pi} K(v,u) f(v)\, \tilde
 o(v)
 .
\ee
This gives again the series \re{intuZOTb}.

 \section{Conclusion}

 We proposed a method for computing the finite volume (or finite
 temperature for the mirror theory) observables in (1+1)-dimensional
 field theories with factorised diagonal scattering and no bound
 states.  The method is based on an exact treatment of the sum over a
 complete set of eigenstates of the Hamiltonian of the mirror theory
 using a graph expansion of the Gaudin measure   using  the
 Matrix-Tree Theorem.  
 The free energy and the observables are expressed in terms of tree Feynman graphs.
 The vertices of such a graph correspond to virtual particles winding multiple times around 
 the compact dimension and the oriented propagators correspond to scattering kernels.
  The method generalises trivially to
 the case of a theory with bound states.  It is very natural to
 conjecture that   the method can be generalised to theories with
 non-diagonal scattering.

  The tree expansion derived here does not use  relativistic invariance, 
  hence the scattering matrix is not necessarily of difference form. 
  Our principal motivation comes from AdS/CFT, where the 
  world sheet (1+1)-dimensional field theory is not Lorentz invariant.
  We  believe that after being generalised for a theory with 
  non-diagonal scattering and bound states,  our construction will 
   help to give  a renormalised  formulation 
  of the hexagon   proposal of \cite{BKV1}  for   computation of  correlation functions
  of trace operators.

Another exercise would be to re-derive the $g$-functions in the case of  integrable 
boundaries   \cite{LeClair:1995uf, Dorey:2004aa}. 
The exact $g$-function for diagonal scattering is known \cite{Pozsgay:2010tv}
but the extension to non-diagonal scattering is still out of reach.
The method might be also relevant for the one-point functions in AdS/dCFT 
\cite{KdLZ-1pf-2015,deLeeuw:2017cop}.

  \section*{Acknowledgments}

 We thank Benjamin Basso for  enlightening  discussions,  to
 Zoltan Bajnok   for bringing to our attention ref. \cite{Woynarovich:2010wt}. 
 This research was supported in part by
 the National Science Foundation under Grant No.  NSF PHY11-25915.

  \bigskip
  \bigskip

\noindent
{\it Note added}
\medskip

\noindent
 After
 the completion   of this work we  learned about  the earlier papers by 
G. Kato and M. Wadati  \cite{2001JMP....42.4883K, 2001PhRvE..63c6106K,  2002JMP....43.5060K, 
2004JPSJ...73.1171K},  where  the expression for the free energy of the  
 Lieb-Liniger model and the XXX Heisenberg ferromagnetic has been obtained 
 by a direct combinatorial method which is essentially  identical to the one we are proposing here.
We thank   Bal\'azs Pozsgay for bringing  these works to our knowledge.

%
%
\providecommand{\href}[2]{#2}\begingroup\raggedright\endgroup

\end{document}